# An Enriched Constitutive Model for Fracture Propagation Analysis using the Material Point Method

Giang D. Nguyen[1, a]

[1]School of Civil, Environmental and Mining Engineering, The University of Adelaide, Australia

[a]g.nguyen@adelaide.edu.au / giang.nguyen@trinity.oxon.org

**Abstract.** We develop a novel constitutive modeling approach for the analysis of fracture propagation in quasi-brittle materials using the Material Point Method. The kinematics of constitutive models is enriched with an additional mode of localized deformation to take into account the strain discontinuity once cracking has occurred. The crack details therefore can be stored at material point level and there is no need to enrich the kinematics of finite elements to capture the localization caused by fracturing processes. This enhancement also removes the drawback of classical smeared crack approach in producing unphysical snapping back constitutive responses when the spatial resolution is not fine enough. All these facilitate the implementation of the new approach in the Material Point Method for analysis of large scale problems. Numerical examples of fracture propagation are used to demonstrate the effectiveness and potentials of the new approach.

**Introduction**

The Material Point Method (MPM; [1, 2]) is an enhanced derivation of the Finite Element Method (FEM) with moving integration points. Loosely speaking it is different from FEM in separating the computational grid with the representation of solids/structures using material points. The material points are allowed to move in and out of their elements. An algorithm is therefore required to map back and forth information from material points to finite element nodes for the solution of equilibrium equations and update of material points' quantities (e.g. velocity, stress, strain). The finite element grid can be reset (or usually kept fixed) after a computational cycle, when all information has been mapped back to material points and their positions updated. The MPM therefore combines the advantages of both Eulerian (fixed grid) and Lagrangian (moving integration points) approaches, including avoidance of mesh distortion under large deformation, and can automatically handle no-slip contacts for impacting bodies.

The representation of solids/structures with material points facilitates the use of any constitutive models and/or failure criteria associated with the material points. Examples for fracture propagation using MPM include smeared crack approach for the analysis of sea ice fracture [2]. This approach to fracture propagation analysis is easy to implement in any MPM code and can handle, at the constitutive level, multiple intersecting cracks. However besides the unphysical scaling of fracture properties, it suffers from the snap back constitutive response if the spatial resolution is not sufficiently fine, due to the lack of enhanced element kinematics to deal with discontinuities.

On the other hand, enhancement to the finite element kinematics to capture discontinuities usually requires adding and storing additional quantities (such as crack sizes and orientations) independently or in the enriched elements. Alternatively, information such as crack sizes and orientation can also be stored at material points and mapped to grid nodes to construct an additional element kinematic field to account for the stress relaxation due to cracking [3]. In the literature, the CRAMP (CRAcks with Material Points [4]) algorithm provides another kind of enhancement in which the kinematics is enhanced via the use of multiple velocity fields, due to the presence of explicitly modeled cracks. This line of approach has obtained great success and been adapted to model a variety of problems involving material cracking ([4, 5]). However the above mentioned approaches may become more complicated and impractical when dealing with complex crack patterns involving many intersecting cracks.

We take a different approach to modeling fracture propagation using the MPM in this paper. Instead of enhancing finite elements to capture the discontinuities caused by cracking, the constitutive model is enriched with an additional mode of localized deformation. In this sense, we want to retain all advantages of the MPM in storing all information at material points, while addressing the strain discontinuity via the enhanced constitutive model. It will also be shown that a length scale will involve in the enrichment, thus providing the constitutive model a good way to capture size effects due to localization. The paper is organized as follows. Enrichment to constitutive models will be presented in the next section, together with a damage model for modeling failure of quasi-brittle materials. This is followed by implementation algorithms for MPM and numerical examples of crack propagations to demonstrate the effectiveness and potentials of the proposed algorithms.

**A kinematically enriched constitutive model**

We take advantage of the fact that the Fracture Process Zone (FPZ) in quasi-brittle failure is usually very small compared to the size of solids/structures under consideration. Instead of embedding an oriented FPZ (or more generally localization zone) in a finite element, we do it in constitutive models. We consider a volume $\Omega$ occupied by a material point depicted in Fig. 1, for a two-dimensional (2D) problem. A localization zone of width $h$ and volume $\Omega_L$ is embedded in this volume. We denote $A$ the projected surface area of $\Omega_L$ on the plane taking $\mathbf{n}$ as its normal vector. It is therefore possible to define an effective size $H=\Omega/A$ of the volume $\Omega$ so that the volume fraction $f$ can be taken as the relative size of the FPZ with respect to $\Omega$:

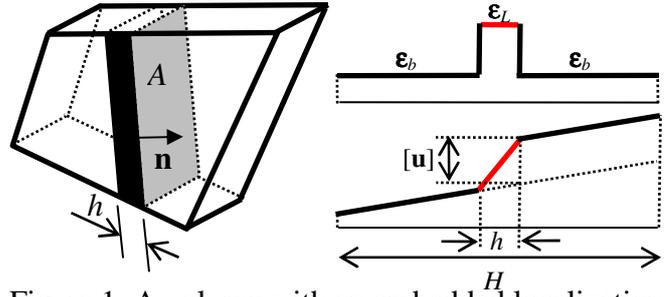

Figure 1: A volume with an embedded localization zone and disp. and strain profiles across the zone

$$f = \frac{\Omega_L}{\Omega} = \frac{h}{H}. \tag{1}$$

The configuration depicted in Fig. 1 can be viewed as a composite material possessing two different phases: elastic phase for the bulk and inelastic phase for the embedded zone. The elastic bulk is assumed here for quasi-brittle failure but is not a compulsory condition for the development. The total strain rate in this case can be written as a volume averaged one with contributions from two different phases:

$$\dot{\boldsymbol{\varepsilon}} = f\dot{\boldsymbol{\varepsilon}}_L + (1-f)\dot{\boldsymbol{\varepsilon}}_b. \tag{2}$$

In the above equation, $\boldsymbol{\varepsilon}$ is the macroscopic strain, $\boldsymbol{\varepsilon}_L$ the strain inside the localization zone, and $\boldsymbol{\varepsilon}_b$ the strain in the bulk continuum. For $h<<H$ the strain rate inside this layer can be approximated as [6. 7]:

$$\dot{\boldsymbol{\varepsilon}}_L = \tfrac{1}{h}\left(\mathbf{n}\otimes[\dot{\mathbf{u}}]\right)^s = \tfrac{1}{2h}\left(\mathbf{n}\otimes[\dot{\mathbf{u}}]+[\dot{\mathbf{u}}]\otimes\mathbf{n}\right). \tag{3}$$

where $[\dot{\mathbf{u}}]$ is relative velocity between opposite sides of the thin FPZ. Due to this inelastic behavior, the stress rate in the bulk continuum is relaxed and can be given by [6, 7]:

$$\dot{\boldsymbol{\sigma}} = \mathbf{a}_b : \dot{\boldsymbol{\varepsilon}}_b = \tfrac{1}{1-f}\mathbf{a}_b : (\dot{\boldsymbol{\varepsilon}} - f\dot{\boldsymbol{\varepsilon}}_L). \tag{4}$$

For elastic behavior in the bulk, $\mathbf{a}_b$ denotes the elastic stiffness tensor. On the other hand, inelastic response is lumped onto the thin band and governed by the following generic constitutive relationship, with $\mathbf{a}_L^T$ denoting the tangent stiffness of the material inside the localization zone:

$$\dot{\boldsymbol{\sigma}}_L = \mathbf{a}_L^T : \dot{\boldsymbol{\varepsilon}}_L. \tag{5}$$

As can be seen, we treat the material as a composite one consisting of two different phases with corresponding behaviors. These behaviors are connected via an internal equilibrium condition to maintain the continuity of traction across the boundary of the thin localization band:

$$(\dot{\boldsymbol{\sigma}} - \dot{\boldsymbol{\sigma}}_L) \cdot \mathbf{n} = 0. \tag{6}$$

The incremental stress-strain relationship of the enriched material model can then be obtained by substituting (3-5) into (6). We then obtain:

$$\tfrac{1}{1-f} \mathbf{a}_b : \left[ \dot{\boldsymbol{\varepsilon}} - \tfrac{f}{h} (\mathbf{n} \otimes [\dot{\mathbf{u}}])^s \right] \cdot \mathbf{n} = \mathbf{a}_L^T : \tfrac{1}{h} (\mathbf{n} \otimes [\dot{\mathbf{u}}])^s \cdot \mathbf{n}. \tag{7}$$

Therefore for a given macroscopic strain rate, the velocity jump $[\dot{\mathbf{u}}]$ can be worked out as:

$$[\dot{\mathbf{u}}] = \mathbf{C}^{-1} \cdot (\mathbf{a}_b : \dot{\boldsymbol{\varepsilon}} \cdot \mathbf{n}). \tag{8}$$

where:

$$\mathbf{C} = \tfrac{1-f}{h} \left( \mathbf{n} \cdot \mathbf{a}_L^T \cdot \mathbf{n} \right) + \tfrac{f}{h} \left( \mathbf{n} \cdot \mathbf{a}_b \cdot \mathbf{n} \right). \tag{9}$$

Substituting (9) into (4) leads to the stress rate in the following form:

$$\dot{\boldsymbol{\sigma}} = \tfrac{1}{1-f} \mathbf{a}_b : \left[ \dot{\boldsymbol{\varepsilon}} - \tfrac{f}{h} \left[ \mathbf{n} \otimes \left( \mathbf{C}^{-1} \cdot (\mathbf{a}_b : \dot{\boldsymbol{\varepsilon}} \cdot \mathbf{n}) \right) \right]^s \right]. \tag{10}$$

The composite response in this case is governed by the behavior of different phases (Eq. 4-5) and their corresponding sizes. In principle, any constitutive relationship can be used for (4) and (5), as the generic algorithm described here requires only the stiffnesses $\mathbf{a}_b$ and $\mathbf{a}_L^T$. Further details can be found in [6, 7]. For quasi-brittle failure, it is reasonable to assume the elastic unloading for the bulk, while a damage model can be used to describe progressing failure inside the localization band. Therefore we take the following damage model governed by the following constitutive relationships [9]:

Stress-strain relationship $\boldsymbol{\sigma}_L = (1-D) \mathbf{a}_L : \boldsymbol{\varepsilon}_L.$ (11)

Damage criterion $y = \dfrac{1}{2(1-D)^2} \left( \boldsymbol{\sigma}_L^+ : \mathbf{a}_L^{-1} : \boldsymbol{\sigma}_L^+ \right) - F(D) \leq 0.$ (12)

where $\mathbf{a}_L$ is the elastic stiffness tensor and $D$ a scalar damage variable. The eigenvalue decomposition [8] is used to decompose the stress tensor $\boldsymbol{\sigma}_L$ into positive ($\boldsymbol{\sigma}_L^+$) and negative parts. Function $F(D)$ in the above expression governs the damage evolution and inelastic response of the material inside the localization zone. It takes the following form [9]:

$$F(D) = \dfrac{f_t'^2}{2E} \left[ \dfrac{E + E_p (1-D)^n}{E(1-D) + E_p (1-D)^n} \right]^2. \tag{13}$$

The above function uses the uniaxial tensile strength $f_t'$, and two parameters $E_p$ and $n$ determined from the fracture energy of the material. Details on how to do that can be found in [9].

**Implementation matters**

The new enriched constitutive structure allows two different behaviors integrated in a constitutive model, together with corresponding sizes. This facilitates the MPM implementation of the approach as a material point now can carry both the elastic bulk and cracking behaviors with an embedded crack (Fig. 2). A continuum damage type one described by Eqs. (11-13) is used in this paper for illustration purpose and for quasi-brittle failure the orientation of the localization band is determined based on the first principal stress. Discrete constitutive models such as the cohesive crack can also be easily integrated in the approach. Details on this have been presented in [7]. Although the memory storage for a material point is double that of a regular constitutive model, due to the presence of an additional stress and strain inside the localization zone, it only applies to cracked material points. We know that the number of cracked material points is usually small compared to the total number of material points in a discretized solid, due to the localized nature of failure.

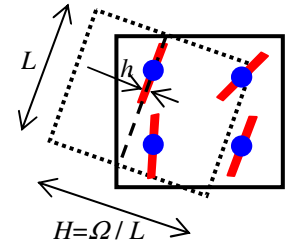

Figure 2: Determination of effective size $H$

The stress return algorithm for the above constitutive structure requires enforcing the traction continuity (6), besides classical stress update algorithms for the inelastic constitutive behavior inside the localization band. It has been described at length in [7]. The focus here is the interface with the MPM, in which each material possess its own size. The effective size $H$ in this case is determined from the element, not the material point, to enforce the reproduction of the correct fracture energy [10]. Following this, an algorithm based on the crack orientation and element geometry, depicted in Fig. 2, is proposed. Alternatively, a simplified one taking $H$ as the square root of element area in 2D also yield satisfactory results, while facilitating the implementation in any existing numerical code [7].

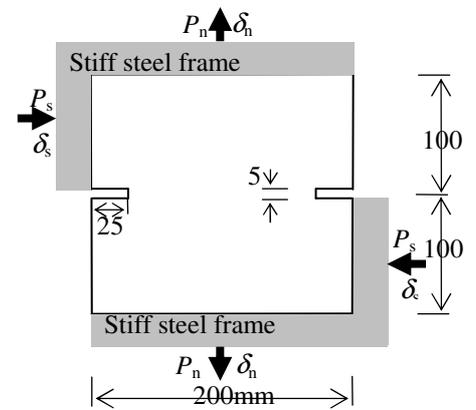

Figure 3: Mixed mode fracture

**Numerical examples**

We use the above enriched model and algorithms for the analysis of quasi-brittle failure in concrete. The mixed mode cracking of a double edge notched (DEN; Fig. 3) specimen is numerically simulated in this example. The experimental tests of the DEN specimens examined were carried out by [11], and the material properties are: Young modulus $E$=32000MPa, Poisson's ratio $\nu$=0.2, uniaxial tensile strength $f_t$'=3.0MPa, fracture energy $G_F$=0.011Nmm/mm$^2$. For the $h$=0.05mm, the calibration procedure described in [9] gives $E_p$=8.184MPa and $n$=0.128. We use three different uniform meshes with finite element sizes of 2.5*2.5mm$^2$ (coarse), 1.25*1.25mm$^2$ (medium) and 0.625*0.625mm$^2$ (fine).

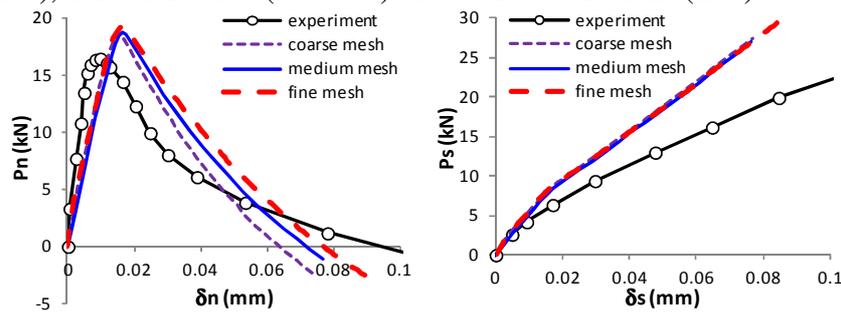

Figure 4: Mesh dependent numerical results.

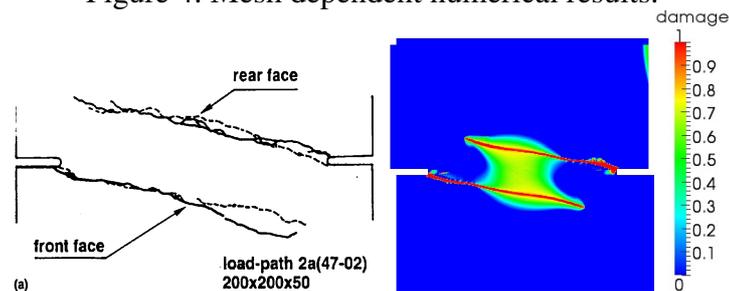

Figure 5: Crack pattern (left: experiment; right: prediction).

The numerical response in Fig. 4 shows the insensitiveness of the results with respect to the resolution of the spatial discretization. The prediction of fracture propagation is compared with experimental observation in Fig. 5, showing a good match in the crack pattern. The predictions can be improved with the use of more advanced constitutive models for the localization zone. However this is not covered within the scope of this paper.

The development of micro-cracks together with the diffuse to localized failure of a Representative Volume Element (RVE) made of cement matrix composite are also simulated here using our in-house MPM code and the approach and model described above. The RVE size is 25mm and thickness 10mm; plane stress condition is assumed and the uniform mesh sizes are 0.25mm and 0.125mm for the coarser and finer meshes, respectively. The generation of sample is facilitated by the use of the MPM and all three phases of the composite material including cement matrix, inclusions and their weak interfacial transition zones (ITZ) can be generated using material points (Fig. 6). We take $h$=0.005mm, and the other fictitious material parameters are listed in the below table.

| Parameters | Inclusion | Matrix | ITZ |
|---|---|---|---|
| $E$ (MPa) | 45000 | 30000 | 30000 |
| $f_t'$ (MPa) | 3.0 | 2.0 | 1.5 |
| $E_p$ (MPa) | 10 | 10 | 20 |
| $n$ | 0.179 | 0.179 | 0.179 |

Table 1: Material parameters for the RVE analysis.

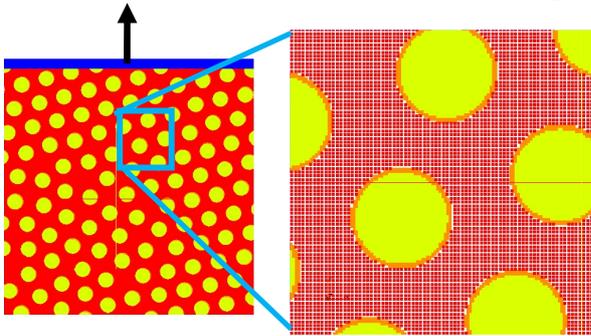

Figure 6: Tension of a RVE made of cement matrix composite.

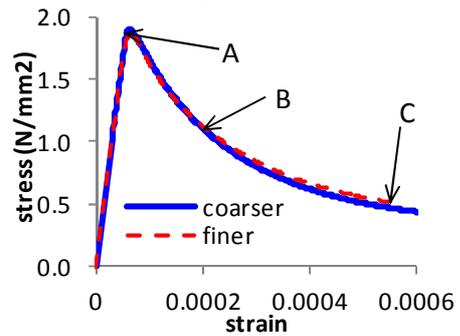

Figure 7: Mesh independent stress strain response.

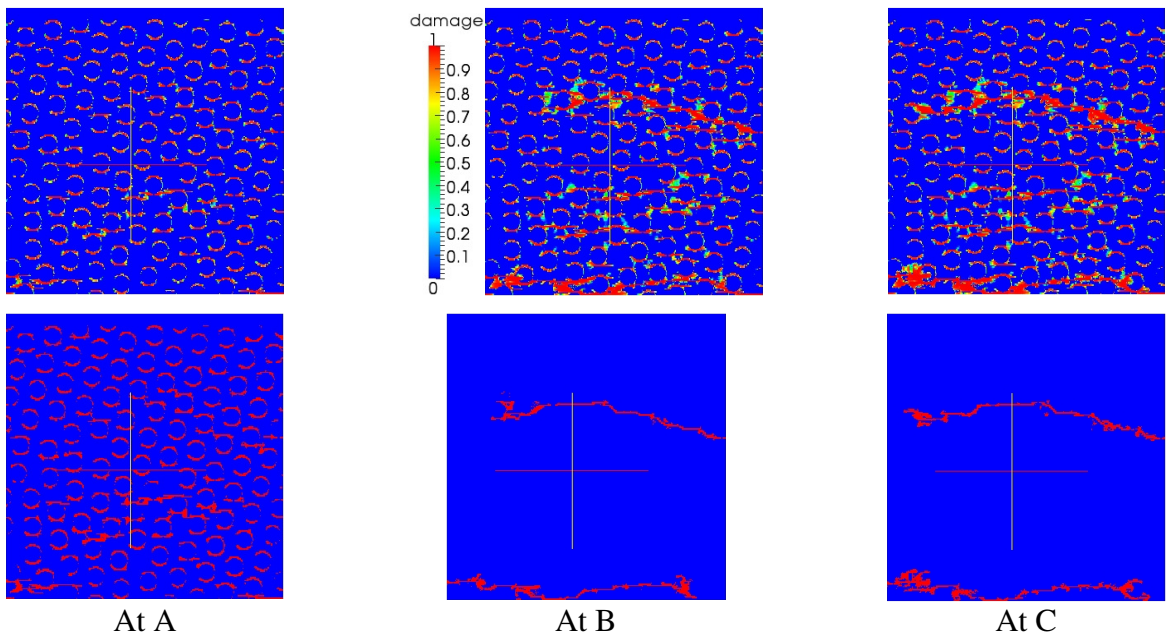

At A        At B        At C

Figure 8: The diffuse to localized failure corresponding to stages in the stress strain curve (Fig. 6) (upper: damage $D$; lower: damage increment $\Delta D$ (only patterns are of concern)).

The sample response is insensitive with respect to the resolution of the discretization (Fig. 7). This again confirms the effectiveness of the proposed approach. There is no need to artificially scale the fracture properties, as in smeared crack approach. The micro-cracking process depicted in Fig. 8 shows the transition from diffuse to localized failure once the peak stress is reached. The total damage indicates crack patterns, while its increment shows the active FPZ (Fig. 8). Crack branching due to the effects of inclusions is also clearly seen. The proposed approach and implementation algorithms worked well in this example (~80000 degrees of freedom). This is a demonstration of the numerical stability and effectiveness of the implementation and our in-house MPM code. Further developments are well on the way for the study of micro fracturing in quasi-brittle materials.

**Conclusions**

An enriched constitutive modeling structure for the fracture propagation analysis using the MPM was developed. The proposed approach embeds an enhanced strain into any constitutive model to help deal with localized mode of deformation. This is the key issue missing in classical constitutive models, preventing them to correctly capture size effects induced by localized failure. The new constitutive modeling structure facilitates the implementation of the approach in any existing numerical code, especially the MPM, when the all details on the fracture are kept at material points, not finite element. The integration of this approach in an MPM code facilitates the micromechanical study of failure in quasi-brittle materials that usually involves complicated micro-cracking patterns.